\documentclass[12pt,preprint]{aastex}
\usepackage{epsfig}
\usepackage{times}
\def\la{\lower.5ex\hbox{$\; \buildrel < \over \sim \;$}}
\def\ga{\lower.5ex\hbox{$\; \buildrel > \over \sim \;$}}
\def\apj{ApJ}

\def\apjl{ApJL}

\def\physrep{Physics Reports}

\shorttitle{Hydrogen 2p--2s signal from recombination and reionization}
\shortauthors{Sethi et al.}

\begin{document}

\title{Hydrogen 2p--2s transition: signals from the \\ 
epochs of recombination and reionization}

\author{Shiv. K. Sethi, Ravi Subrahmanyan, D. Anish Roshi} 
\affil{Raman Research Institute , Sadashivanagar, Bangalore 560080, India \\ email: sethi@rri.res.in, rsubrahm@rri.res.in, anish@rri.res.in}

\begin{abstract}
We propose a method to study the  epoch of reionization  based
on the possible observation of 2p--2s fine structure lines from the neutral
hydrogen outside the cosmological H~{\sc ii} regions enveloping QSOs and other
ionizing sources in the reionization era. We show that for parameters typical of
luminous sources observed at $z \simeq 6.3$
the strength of this signal, which is proportional to the H~{\sc i} fraction,
has a  brightness temperature $\simeq 20 \, \mu K$ for a fully neutral medium. 
The fine structure line from this redshift is  observable at $\nu \simeq 1 \, \rm
GHz$ and we  discuss prospects for the detection with several operational and future radio
telescopes.  We also compute the characteristics of this signal from the epoch of 
recombination: the peak brightness is expected to be $\simeq 100 \, \mu K$; this signal appears
 in the frequency range 5-10 MHz.  The signal from the recombination era is
 nearly impossible to detect owing to the extreme brightness of the
 Galactic emission at these frequencies.
\end{abstract}


\section{Introduction}

Even though the existence of Hydrogen fine structure lines and their explanation 
using Dirac's atomic theory has been known for close to a century,
 a Hydrogen fine structure 
line has never been detected from an astrophysical object. 
An  interesting Hydrogen fine structure line is the $2p\hbox{--}2s$ transition.
 The main difficulty in detecting this line is that the line strength
is proportional to the population of either the $2p$ or $2s$ states which, being
excited states, are not so readily populated in most astrophysical circumstances.
Moreover, the line width of the excited $2p$ state, which is determined by its decay time,
is large (99.8~MHz), making the detection of the fine structure line a difficult observation.
One  astrophysical setting where the feasibility of detecting such a line has been studied 
is the interstellar H~{\sc ii} regions (see, {\it e.g.}, Dennison, Turner, \& Minter 2005
and references therein; Ershov 1987); in H~{\sc ii} regions, the excited levels are 
populated by recombination. 

Here we consider two cosmological settings in which
the excited levels are populated by either recombination or 
pumping by Lyman-$\alpha$ photons from an external source:
(a) {\it The Recombination epoch}: 
The Universe makes a transition from a fully ionized to an almost fully neutral medium at 
$z \simeq 1089$ (Spergel et~al. 2006; for details see, {\it e.g.}, Peebles 1993 and 
references therein).  During this era, as the density and temperature of the 
Universe drops, recombination is stalled owing to a high Lyman-$\alpha$ radiation density 
and progresses either by the depopulation of the $2p$ state owing to redshifting 
of the photons out of the line width or the 2-photon decay of the $2s$ state. 
This results in a significant $2s$ and $2p$ level population during the recombination era. 
(b) {\it The Reionization epoch}:
Recent observations suggest that the  universe made a  transition from nearly fully neutral to 
fully ionized within the redshift range $6 \la z \la 15$ (Page  et~al. 2006;
White et~al. 2003; Fan et~al. 2002; Djorgovski et~al. 2001; Becker et~al. 2001). It is widely
believed that this `reionization' was achieved by the percolation of 
individual H~{\sc ii} regions around the sources of reionization. The nature of 
these sources is not well understood: 
they might be Pop~{\sc iii} stars, active galactic nuclei or star-forming galaxies.
During this epoch, a signal from the $2p-2s$ fine-structure transition might originate 
from either within the cosmological H~{\sc ii} regions or from
the almost fully neutral medium surrounding the H~{\sc ii} region. The 
level population of the first excited state in the former case would be 
largely determined by recombinations and in the latter case 
by Lyman-$\alpha$ photons from the central source.  We shall show below that 
for 
most cases of interest the fine-structure line from within the cosmological H~{\sc ii} 
region might be negligible as compared to the signal from the 
regions immediately surrounding the H~{\sc ii} region. 

Throughout this work we adopt the currently-favoured $\Lambda$CDM model: spatially flat
with $\Omega_m = 0.3$ and $\Omega_\Lambda = 0.7$ 
(Spergel et al. 2006; Riess et~al. 2004; Perlmutter et al. 1999) with  
$\Omega_b h^2 = 0.022$ (Spergel et al. 2006; Tytler et al. 2000) and
$h = 0.7$ (Freedman  et al. 2001).

\section{Fine-structure lines from the reionization epoch}
Subsequent to the recombination of the primeval baryon gas at 
redshift $z\simeq 1089$ (Spergel  et~al. 2006)
and the transformation of the gas to an almost completely neutral state, it is
believed that the gas was reionized during epochs corresponding to the redshift range
$6 \la z \la 15$.  WMAP measurements of cosmic microwave background radiation
(CMBR) anisotropy in total intensity and polarization
have been used to infer that the baryons were likely neutral at redshifts $z \ga 12-15$; however,
the detection of CMB polarization anisotropy requires substantial 
ionization by about $z \simeq  11$ (Page et~al. 2006).
Observationally, the Gunn-Peterson (GP) test shows
that the universe is highly ionized at redshifts lower than $z \simeq 5.5$; the 
detection of GP absorption at greater redshifts suggests that the neutral fraction
of the intergalactic hydrogen gas rises to at least $10^{-3}$ in the redshift range 
$5.5 \la z \la 6$, and that reionization was not complete till about $z \simeq  6$ 
(White et~al. 2003; Fan et~al. 2002; Djorgovski et~al. 2001; Becker et~al. 2001).  
However, from the GP test alone it is not possible to infer the 
neutral fraction of the medium; it only gives a rather weak lower 
limit of $\simeq 10^{-3}$ on the neutral fraction of the 
universe for $z \ga 6$.  From other considerations it is possible to put
more stringent bounds on the neutral fraction; for example, Wyithe \& Loeb 
(2004) obtain a lower 
limit of $0.1$ on the neutral fraction of the universe at $z \simeq 6.3$ 
(see also Mesinger \& Haiman 2004). 

Our understanding of the nature of the sources that caused the reionization is
far from complete.  The transition from an almost completely neutral gas 
to a highly ionized gas during redshifts  
$6 \la z \la 15$ is a key problem in modern cosmology and considerable theoretical
and experimental efforts are currently directed at this unsolved problem.
Here we propose a new method, based on the $2p-2s$ fine-structure transition, 
to determine the evolution of the neutral fraction of the intergalactic medium 
within this epoch. 

Owing to fine structure  splitting, the two possible transitions between 
the $2s$ and the $2p$ states are: $2p_{1/2}\hbox{--}2s_{1/2}$ at a frequency 
$\simeq 1058 \, \rm MHz$ that has an Einstein $A$ coefficient $1.6 \times 10^{-9} \, \rm s^{-1}$ 
and $2p_{3/2}\hbox{--}2s_{1/2}$ at a frequency  $\nu_{ps} \simeq  9911 \, \rm MHz$
that has an Einstein $A$ 
coefficient $8.78 \times 10^{-7} \, \rm s^{-1}$.
The Einstein $A$ coefficient for the latter transition is more than an order of 
magnitude greater than the former; therefore, in this work we consider 
only the $2p_{3/2}\hbox{--}2s_{1/2}$ transition and hereinafter we refer to this
specific transition simply as the $2p\hbox{--}2s$ transition.

The ionizing UV photons from  sources in the 
reionization era create `Stromgren spheres'. Whereas
the gas in the cosmological H~{\sc ii} regions are highly ionized by the photons, 
the ionization level of the  gas beyond the Stromgren spheres is
determined by the history of the gas, the density, and the mean specific intensity of the 
background ionizing photons, which 
includes both the photons diffusing out of the Stromgren spheres as well as 
the background radiation field.  In this work we assume that the ionizing
sources at these high redshifts are AGNs and in illustrative examples adopt
parameters of a few QSOs that have been observed at $z \simeq 6$.    
The photons at the Lyman-$\alpha$ transition frequency (here and throughout, 
unless otherwise specified, 
we shall continue to refer to frequencies between Lyman-$\alpha$ and
 Lyman-limit as 'Lyman-$\alpha$) 
from a high redshift QSO escape the mostly-ionized Stromgren sphere and are strongly
scattered and absorbed in the medium beyond. The population of the 
$2p$ level in this region is determined  by (a)
the intensity of Lyman-$\alpha$ photons from the central source, (b) 
recombination rate of free electrons, (c)  absorption of CMBR photons 
by electrons in the $2s$ state (it is assumed here and throughout this work that the only 
radio source at high redshifts is the CMBR) and (d) collisional transition 
from the $2s$ state.  The $2s$ state
is populated via (a) the recombination rate of free electrons, (b)  collisional transfer 
of atoms from the $2p$ state, (c) the spontaneous  decay of the $2p$ state, and
(d) transition from the $2p$ state stimulated by CMBR photons.  
Additionally, the absorption of photons 
from the central sources, with energy equal to or in excess of the Lyman-$\beta$  transition, 
would result in electronic transitions 
to the second excited state, which could be followed by spontaneous decay to the $2s$ state.
(It might be pointed out here that both the $2s$ and $2p$ states could also be 
populated by atoms cascading from excited states with  $n > 3$. In particular,
all photons absorbed from $1s$ states to any excited state can directly
de-excite to the $2s$ level. However, the rate of transition from $1s$ to 
any excited state is roughly $\propto 1/n^3$ (e.g. Rybicki \& Lightman 1979)
and, therefore, we include only the most dominant transition in each case.)

The population of the ground state is denoted by $n_{1s}$.  
We denote the level populations of the the two states---$2s_{1/2}$ and $2p_{3/2}$---by the number
densities $n_{2s}$ and $n_{2p}$; these may be solved for, respectively, 
from the following two equations of detailed balance:
\begin{eqnarray}
f \alpha_{\scriptscriptstyle \rm B} n_i^2 + c B_{2p2s}n_{2p} n_{\scriptscriptstyle \rm CMBR}+ 
C_{\rm ps}n_i n_{2p} & + & A_{2p2s} n_{2p} + 
c n_{1s}p_{32}  \int B_{13,\beta} \phi_{13}(\nu) n_{\alpha}(\nu) d \nu \nonumber \\
& = &  A_{2s1s} n_{2s} + C_{sp}n_i n_{2s} +cB_{2s2p}n_{2s} n_{\scriptscriptstyle \rm CMBR},
\label{levpops}
\end{eqnarray}
\begin{eqnarray}
(1-2f) \alpha_{\scriptscriptstyle \rm B} n_i^2 +  c B_{2s2p}n_{2s} 
n_{\scriptscriptstyle \rm CMBR}& + &  C_{\rm sp}n_i n_{2s} + c n_{1s} 
\int B_{12,\alpha} \phi_{12}(\nu) n_{\alpha}(\nu) d \nu \nonumber \\
& = &  A_{2p1s} n_{2p} + C_{ps}n_i n_{2p} +  cB_{2p2s}n_{2p} n_{\scriptscriptstyle \rm CMBR}.
\label{levpopp}
\end{eqnarray}
Here $f$ is the fraction of all the atoms that recombine to the $2s$ state. 
In equilibrium $f = 1/3$ as the $n=2$ state splits into three doublets:
$2p_{1/2}$, $2p_{3/2}$ and $2s_{1/2}$.  $\alpha_B$ 
is the recombination coefficient and $n_i$ is the density of the ionized gas. 
$B_{2p2s} = B_{2s2p} = c^2/(8\pi\nu_{ps}^3) A_{2p2s}$ is 
the Einstein $B$ coefficient for the $2p_{3/2}\hbox{--}2s_{1/2}$ transition in terms of the 
corresponding Einstein $A$ coefficient $A_{2p2s}$ (note that the 
two $B$ coefficients are equal as the two states 
have the same degeneracy). $C_{ps} = C_{sp} = 5.31 \times 10^{-4} \, \rm cm^3 \, s^{-1}$  is the rate coefficient of transition owing to collisions
with electrons.  $n_{\scriptscriptstyle \rm CMBR}$ is the number density of CMBR photons 
within the transition line width. 
$n_\alpha(\nu)$ is the number density (per unit frequency) of 
photons with frequency equal to or larger than  the Lyman-$\alpha$  
frequency (and smaller than the Lyman-limit frequency) 
at any location; $\phi_{13}$ and $\phi_{12}$ are, respectively, the line profiles corresponding 
to the Lyman-$\beta$ and Lyman-$\alpha$ transitions, and $p_{32}$ is the probability for the 
electron transition to the $2s$ state following excitation to $n=3$ via absorption of a
Lyman-$\beta$ photon.  We have not included the 
induced Lyman-$\alpha$ transition because  the number 
density of atoms in the $2p$ state is negligible as compared to that 
in the $1s$ state. 
$p_{32}$ is the probability that an atom in the third excited state ($3p$) will
decay to the $2s$ state.  $A_{2s1s}$ is the Einstein $A$ coefficient corresponding
to the 2-photon decay of the $2s$ state. Other symbols have their usual meanings. 

Owing to the fact that the mostly neutral medium in the vicinity of the 
cosmological Stromgren spheres is optically thick to Lyman-$\alpha$ 
scattering, the Lyman-$\alpha$ photons from the decay of the $2p$ state 
are strongly scattered by the gas. Therefore, $n_\alpha(\nu)$ will contain contributions
from both the Lyman-$\alpha$ photons from the central source
as well as the Lyman-$\alpha$ photons that arise from recombinations outside the 
Stromgren sphere and are multiply-scattered therein: 
$n_\alpha(\nu) =  n_\alpha^{\rm source}(\nu) + n_\alpha^{\rm rec}(\nu)$. 
We neglect the multiply-scattered Lyman-$\alpha$ photons 
from the Stromgren sphere that have been reprocessed via recombination within the
Stromgren sphere because these would be redshifted redward of the Lyman-$\alpha$ line before
encountering the boundary of the Stromgren sphere.
The scattering of recombination photons in an optically thick, expanding medium 
is a complex problem (Field 1959; Rybicki  \& Dell'Antonio 1994).
One of its first applications was to study the recombination of primeval 
plasma (Peebles 1968;  Zeldovich, Kurt \& Sunyaev 1969). In these analyzes it 
was implicitly assumed that apart from 2-photon decay, in an expanding universe
the dominant effect that results in resonant photons ceasing interaction with the gas, and
leaving the system, is its redshifting out of the line profile.  
The effect of scattering off the moving 
atoms was deemed to be either negligible or at best comparable. This 
assumption has been borne out by more recent detailed analysis that have taken into
account the effect of scattering on the photon escape (Krolik 1990).
Taking only the redshift as the main agent  of photon escape, it can be 
shown that the net effect of the scattering of a resonance photon before
it drops out of consideration is to reduce the decay time of the $2p$ state
from $A_{21} = 6.2 \times 10^8 \, \rm s^{-1}$ to $A_{21}/\tau_{\scriptscriptstyle \rm GP}$ 
(Zeldovich  et~al. 1969; for more recent work see, {\it e.g.}, 
Chluba, Rubino-Martin  \&  Sunyaev 2007, Seager, Sasselov \& Scott 1999 
and references therein; we give a concise derivation in Appendix~A).  In this
expression, the Gunn-Peterson
optical depth $\tau_{\scriptscriptstyle \rm GP} = [3/(8\pi H)]A_{2p1s}\lambda_\alpha^3 n_{1s}$.  

Similar complications exist in computing $p_{32}$, the probability 
that an atom in the third excited state will decay spontaneously to the $2s$ state, in an
optically thick medium. In optically thin media, $p_{32} = A_{32}/(A_{32} + A_{31})$;
on the other hand, in an optically thick medium, we 
would need to take into account the `trapping' of the 
Lyman-$\beta$ photon owing to resonant scattering. The effect of this 
scattering 
in an optically thick medium  would be
that the fraction of photons that decay directly to the ground state are reabsorbed 
`locally' to the third excited state and, therefore, all photons absorbed to 
the third 
excited state result in  an H$\alpha$ photon and an atom in the $2s$ state.
 This means that
the appropriate value of $p_{32}$ is close to unity in an optically thick
 medium:
in this work we assume $p_{32} = 1$.  

The astrophysical setting in which we seek solutions to the 
algebraic equations above is cosmological H~{\sc ii} regions at high redshift. In 
particular, we are interested in the signal from the neutral region surrounding
the cosmological H~{\sc ii} region.  For a fully neutral inter-galactic medium
(IGM) at $z \simeq 6.5$,
$\tau_{\scriptscriptstyle \rm GP} \simeq 6 \times 10^5$.
If we adopt spectral luminosities
corresponding to QSOs observed at these high redshifts, it may be shown that  
the populating of the $2p$ state via direct recombinations from the 
free-free state, 
collisional transfer from the $2s$ state, and upward transitions from 
the $2s$ to $2p$ 
state arising from absorption of background CMBR photons may all be neglected.
(The relevant parameters  are:  $n_i \simeq n_b \simeq  2.8(1+z)^3 \, \rm cm^{-3}$
in the H~{\sc ii} region
surrounding the sources, with $n_i$ expected to be much smaller in the
neighboring mostly neutral medium; the number density of CMBR photons 
that might cause a $2p\hbox{--}2s$ transition is $ \simeq 5(1+z) \, \rm
cm^{-3}$; the number density of Lyman-$\alpha$ photons
from the central source, assuming luminosities typical of SDSS quasars at $z
\simeq 6.5$ (more
details in \S 4), is $n_\alpha \simeq 10^{-4} \, \rm cm^{-3}$. 
First, for these parameters, the dominant process that populates the excited state
is the pumping by Lyman-$\alpha$ photons. Second, it may be readily
verified that for these plausible values for the parameters 
the signal expected from the H~{\sc ii} region surrounding 
the central source is negligible as compared to the signal
from the surrounding neutral region.)
Given that the Lyman-$\alpha$ flux from the central QSO
is the dominant causative factor for populating the $2p$ state,  
the two equations (\ref{levpops} and \ref{levpopp}) that determine the level populations
is  simplified.  The number density of atoms in the $2p$ state is given
approximately by:
\begin{equation}
n_{\rm 2p} 
\simeq n_{1s} \Gamma_{\alpha} \tau_{\scriptscriptstyle \rm GP}/A_{2p1s},
\label{num2p}
\end{equation} 
where $n_{\rm 1s} = f_{\rm neu} n_{\rm b}$, with $f_{\rm neu}$ denoting the neutral fraction and
$n_b \simeq 2.7\times 10^{-7} (1+z)^3$~cm$^{-3}$ is the number density of baryons in the 
IGM. $\Gamma_{\alpha} = \int B_{2p1s} \phi_{12}(\nu) n_{\alpha}^{\rm source}(\nu) $ (in units of 
$\rm s^{-1}$) is the transition rate to the $2p$ state owing to the Lyman-$\alpha$
photons from the central source, where $n_{\alpha}^{\rm source}(\nu)$ is the number 
density of Lyman-$\alpha$ photons from the central source alone. 

Similarly, the dominant process that determines the population of the $2s$ state is the absorption
of Lyman-$\beta$ photons (for details, see the discussion above
and \S 4)  and the subsequent decay to the $2s$ state. 
The number density of atoms in the $2s$ state is:
\begin{equation}
n_{\rm 2s} 
\simeq n_{1s} \Gamma_\beta/A_{2s1s},
\label{num2s}
\end{equation} 
where $\Gamma_\beta  = \int B_{3p1s} \phi_{13}(\nu) n_{\alpha}^{\rm source}(\nu) $ 
(in units of $\rm s^{-1}$) is the transition rate to the $2s$ state owing to the Lyman-$\beta$
photons from the central source. 

\section{Fine structure lines from the epoch of recombination}
The universe made a transition from fully ionized to nearly fully 
neutral at $z \simeq 1089$ (Spergel  et~al. 2006; Peebles 1968; Zeldovich  et~al. 1969). 
This transition is mainly accomplished by the 2-photon decay of the 
$2s$ state and the slow redshifting of the Lyman-$\alpha$ photons which deplete the $2p$ state 
(see Peebles 1993 and references therein for a detailed discussion). 
The Saha ionization formula, valid for thermodynamic equilibrium conditions 
between the hydrogen level populations and free electrons, is a poor approximation for studying 
the epoch of recombination. 
A good approximation for  studying this transition is to assume that all states,
excepting $1s$, are in equilibrium with the CMBR (matter temperature to 
a very good approximation remains equal to the CMBR temperature throughout
this transition) (Seager et~al. 1999; Peebles 1968). However, in this 
approximation, where the matter temperature and all transitions, excepting the Lyman-$\alpha$
line, are in thermal equilibrium, the $2p\hbox{--}2s$ signal is unobservable because the
excitation temperature for this transition equals the background radiation temperature. 
Even though this second approximation might be useful 
for studying the evolution in ionization fraction,
it will be strictly true only if the dominant mechanisms that determine
the level populations of the $2p$ and the $2s$ states are  either interaction 
with the CMBR photons or collisions between 
atoms. As there are  a variety of other processes relevant to the determination of the
level populations---for example, the free decay of either of the two states---a small deviation
from equilibrium is expected in the $2s$ and $2p$ level populations and it 
is our aim here to compute it. 

One approach to this problem is to simultaneously solve for the level populations of 
the $2p$ and the $2s$ states as well as the change in the ionization fraction. 
However, assuming thermal equilibrium between these two states
is a good approximation for solving the evolution of ionization. Therefore, the approach we have
adopted is to solve for the evolution of ionization  using the method of Peebles (1968), 
and use the resulting ionized/neutral fraction to solve for the populations 
of the $2s$ and the $2p$ states using detailed balance.  The resulting equations are:
\begin{eqnarray}
f \alpha_{\scriptscriptstyle \rm B} n_i^2 +  c B_{2p2s}n_{2p} n_{\scriptscriptstyle \rm CMBR}  + 
C_{\rm ps}n_i n_{2p}& + &  A_{2p2s} n_{2p} + 
n_{1s} A_{2s1s}\exp(-(B_1-B_2)/(k_{\rm B}T_{\rm CMBR}))  \nonumber \\
 & = &  A_{2s1s} n_{2s}  +  C_{sp}n_i n_{2s} +  cB_{2s2p}n_{2s} n_{\scriptscriptstyle \rm CMBR} + 
\beta_c n_{2s}
\label{levpop0}
\end{eqnarray}
and
\begin{eqnarray}
(1-2f) \alpha_{\scriptscriptstyle \rm B} n_i^2 +  c B_{2s2p}n_{2s} n_{\scriptscriptstyle \rm CMBR}& + &  
C_{\rm sp}n_i n_{2s}  \nonumber \\
 & = & A_{2p1s} n_{2p}/\tau_{\scriptscriptstyle GP} + 
C_{ps}n_i n_{2p}  +  cB_{2p2s}n_{2p} n_{\scriptscriptstyle \rm CMBR} +\beta_c n_{2p},
\label{levpop1}
\end{eqnarray}
where $B_1 = 13.6 \, \rm eV$ and $B_2 = 2.4 \, \rm eV$ 
are, respectively, the ionization potentials of the ground and the first excited states,
$k_{\rm B}$ is the Boltzmann's constant, and $T_{\rm CMBR}$ is the temperature of the CMBR.
$\beta_c$ is the rate at which  the CMBR photons cause a bound-free transition of electrons from the 
$n=2$ state ($2s$ or $2p$ states).
The various other terms in these equations have the same meanings as in the previous case. The main difference is that
the CMBR photons and baryonic matter at the time of recombination are hot and dense enough to directly
affect the level populations of the excited states by ionizing the 
excited state, and the two-photon capture to the excited state is not 
completely negligible (see Peebles 1993 and 1968 for details of the different
physical processes that are relevant at this epoch).  

\section{Expected brightness of the fine structure lines}
The brightness temperature in the $2p\hbox{--}2s$ transition is:
\begin{equation}
\Delta T_b \equiv T_b - T_{\rm CMBR} = g_\star
{n_{2p}(0)h_{\rm p} A_{\rm 2p2s} \lambda_e^2(1+z)^2 c  \over 8\pi k_{\scriptscriptstyle \rm B}  H(z)}
\left (1 - T_{\rm CMBR}/T_{ex} \right ) \equiv \tau_{ex}(T_{ex}-T_{\rm CMBR}) 
\label{bri_gen}
\end{equation}
Here $n_{2p}(0) = n_{2p}/(1+z)^3$ and $\lambda_e$ is the rest wavelength of the $2p\hbox{--}2s$ transition. 
$T_{ex} = (h_{\rm p}\nu_e/k_{\scriptscriptstyle B})[n_{2p}/(n_{2s} - n_{2p})]$ 
is the excitation temperature corresponding to the transition, where $h_{\rm p}$ is the Planck constant.
$\tau_{ex}$ is the optical depth of the source in the $2p\hbox{--}2s$ signal. 
$g_\star$ takes into account the selection rules for transitions between the $2p_{3/2}$ and $2s_{1/2}$ levels. 
Given the selection rules there are three allowed transitions between these
two states; they occur at frequencies 
$\nu \simeq 9852$, 9875, and 10029~MHz (see, {\it e.g.}, Ershov 1987). 
The first two transitions are blended by the natural width of the 
line, which is approximately 100~MHz, but the third should be observable as a distinct line. 
This implies that $g_\star = 2/3$ if the observing frequency is $\simeq 9900/(1+z)$~MHz. 

\subsection{Expectations for signals from cosmological H~{\sc ii} regions}

For computing the strength of this signal in a typical case, we 
adopt observed parameters of QSO SDSS J1030{+}0524 (see, {\it e.g.}, Wyithe \& Loeb 2004 and references therein),
which is at redshift $z = 6.28$ and shows no detectable flux beyond the QSO Stromgren sphere: the
radius of the Stromgren sphere has been estimated to be $R \simeq 4.5 \,\rm Mpc$ (Mesinger \& Haiman (2004) argue that the size of the Stromgren sphere 
could be roughly 30\% higher; this makes no essential difference to our
results).
It may be noted here that the QSO SDSS J1148{+}5251 also has similar 
parameters.  We arrive at an estimate of 
$n_{\alpha}^{\rm source}(\nu)$ by assuming that $L_{\alpha}$ photons 
per second, with wavelengths corresponding 
to the Lyman-$\alpha$ transition, are emitted by the central source in
 an effective 
frequency range $\Delta \nu_{\rm source}$, and that the photons are 
absorbed at a radial 
distance $R$.  This leads to:
\begin{equation}
n_{\alpha}^{\rm source} = L_{\alpha}/(4\pi R^2 c)\times {1\over \Delta \nu_{\rm source}}.
\end{equation}
We assume typical values: $L_{\alpha} = 10^{58} \, \rm s^{-1}$, $R = 4.5 \, \rm Mpc$ and 
$\Delta \nu_{\rm source} = 30,000 \, \rm km \, s^{-1}$. These lead to the
following estimate for 
the transition rate, $\Gamma_{\alpha}$, in the gas at the boundary of the Stromgren sphere 
arising due to the photons from the central QSO:
\begin{equation}
\Gamma_{\alpha}  \simeq 8 \times 10^{-10}~\rm s^{-1}.
\end{equation}
It should be noted that the mean specific intensity of 
Lyman-$\alpha$ photons in the IGM would also give a non-zero signal.  
Assuming a mean specific intensity of $\simeq 10^{-21} \, \rm erg \, cm^{-2} \, sec^{-1} \, Hz^{-1} \, sr^{-1}$ (this might be needed to couple the HI  spin
temperature to the matter temperature; see e.g. Madau, Meiksin, \&  Rees 1997), the expected signal is many orders of magnitude smaller than we have
computed  from
the outskirt of bright sources. 
If we assume that the central sources are continuum emitters in the 
frequency range between Lyman-$\alpha$ and Lyman-limit frequencies, 
we may assume similar parameter values for computing the expectations for 
$\Gamma_\beta$. 
In that case, 
\begin{equation}
\Gamma_\beta  \simeq \Gamma_{\alpha} {f_{\beta} \over f_{\alpha}},
\label{pho_ratio}
\end{equation}
where $f_\beta$ and $f_\alpha$ are the oscillator strengths of the Lyman-$\beta$  and 
Lyman-$\alpha$ transitions, respectively. From Eqs.~(\ref{num2p}),~(\ref{num2s}) and~(\ref{pho_ratio}):
\begin{equation}
{n_{\rm 2s} \over n_{\rm 2p}} \simeq {f_{\beta} \over f_{\alpha}}{A_{2p1s} \over A_{2s1s} \tau_{\rm \scriptscriptstyle GP}}.
\end{equation}
For a completely neutral medium at $z\simeq 6.4$, $\tau_{\rm \scriptscriptstyle GP} \simeq 6 \times 10^5$, which may be 
used to show that $n_{\rm 2s} \gg  n_{\rm 2p}$ for the parameters of IGM in
the redshift range of interest: $6.4 \la z \la 10$. 
This implies that in the outskirts of the Stromgren sphere surrounding the QSO, 
the transition is expected to be observable as an absorption feature against the CMBR. 
Using Eqs.~(\ref{num2p})~and~(\ref{num2s}) in Eq.~(\ref{bri_gen}), the observable brightness temperature is estimated to be:
\begin{equation}
\Delta T_B \simeq -20~\mu K \left ({f_{\rm neu} \over 1} \right ),
\label{bri_tem}
\end{equation}
where $f_{\rm neu}$ is the neutral fraction of hydrogen outside the Stromgren
sphere, and might  be close to unity.

We have adopted parameters typical of QSOs observed at the edge of the reionization epoch
in estimating the above temperature decrement.  The main
uncertainty above is in the estimation of the `Lyman-$\alpha$' flux from the central source, and
as the observed temperature decrement is directly proportional to this 
flux from the central source, this 
constitutes a major uncertainty in reliably computing the expectations for the signal. 
For QSOs that have strong Lyman-$\alpha$ and Lyman-$\beta$ lines, the blueward side of 
the lines will be strongly absorbed in the medium just beyond the Stromgren sphere
(and this has been observed to happen in many cases),
provided that the blueward side photons have not been
redshifted to frequencies smaller than the Lyman-$\alpha$ frequency while transversing the 
Stromgren sphere. In the case of  QSOs that have large line fluxes and small Stromgren spheres, the 
Lyman-$\alpha$ luminosity $L_{\alpha}$ might be underestimated.  

\subsection{Expectations for the fine-structure line from recombination}
Using equations~(\ref{levpop0}) and (\ref{levpop1}) the level populations 
of the $2p$ and the $2s$ states may be computed.
Solving for the level populations and using equation~(\ref{bri_gen}), we have
computed the expected signal from the 
recombination epoch; the expected signal is shown in Fig.~1. The fine
structure line transition is 
expected to be an absorption feature, with a maximum temperature decrement 
of order 100~$\mu$K at an observing  frequency of about 7~MHz. 
The width of the decrement in frequency space corresponds to a redshift span $\Delta z \simeq 200$, which is 
roughly the width of the visibility function at recombination. 

\section{Prospects for the detection of the cosmological fine structure lines}
As Eq.~(\ref{bri_tem}) shows, a detection of the fine structure line in the outer
regions of cosmological H~{\sc ii} regions is potentially a probe of the cosmological 
neutral hydrogen density in the vicinity of QSOs in the reionization epoch, and might be a tool
for the investigation of the evolution in the neutral fraction with cosmic epoch through
the reionization era. Given the cosmological importance of such a measurement,
and the fact that there does not exist many reliable
methods for the detection of H~{\sc i} at high redshifts (see, {\it e.g.}, Barkana \& Loeb 2001), 
the detection of the $2p\hbox{--}2s$ line transition in the cosmological context 
assumes additional importance. 

In deriving equation~(\ref{bri_gen}) the Hubble expansion was assumed to be the 
only cause for the velocity width in the observed line. However, an important contribution
to the velocity dispersion in the line in this case is the natural width of the 
fine structure line: owing to the rapid decay of the $2p$ state, the natural Lorentzian 
width in the rest frame of the gas is $V_{\rm Lor} \simeq 100 \, \rm MHz$
(see, {\it e.g.}, Dennison et al. 2005).  Therefore, the peak brightness in the 
observed line profile might be suppressed by a factor $V_{\rm exp}/V_{\rm Lor}$, where
$V_{\rm exp}$ is the line-of-sight peculiar velocity dispersion owing to the Hubble flow
across  the region being observed.  However, in the case  $V_{\rm exp} \ga V_{\rm Lor}$ 
this suppression has a negligible effect.  For QSOs at redshift $z \approx 6.5$, the natural 
Lorentzian width of the fine structure line is equivalent to the peculiar
Hubble flow across a proper line-of-sight distance of $\simeq 4 \, \rm Mpc$.
This distance is approximately the size of the Stromgren
spheres around QSOs at that redshift; therefore, the natural width of the line 
does not significantly diminish the expectations, given by
equation~(\ref{bri_gen}), for the peak brightness
temperature.  A second inference is that in
the case of a QSO that has a smaller  Stromgren sphere,
the increase  in mean  brightness  temperature  is roughly proportion to 
the inverse of the radius of the Stromgren sphere: $1/R$,  and not $1/R^2$. 

Another assumption that was made while deriving Eq.~(\ref{bri_gen}) is that the only radio frequency radiation 
that needs to be considered for the determination of the level populations and
brightness temperature decrement is the CMBR. 
It may be that the central ionizing source, which may be a QSO, is radio loud.
There may also exist radio sources  behind the observed Stromgren sphere and within
the  angular region over which  the Lyman-$\alpha$ flux from the 
QSO is  appreciable. Equation (\ref{bri_gen}) may be modified, to account for this, by
replacing $T_{\rm CMBR}$ with $T_{\rm CMBR} + T_{\rm B}$, where $T_{\rm B}$ is the brightness
temperature of the radio source at wavelengths corresponding to the rest frequency of the 
transition, which is $\simeq 9 \, \rm GHz$. The result would be to enhance the 
brightness temperature of the line signal by $\simeq T_{\rm B}/T_{\rm CMBR}$. 
If the radio source is unresolved, it is appropriate to estimate the expected signal 
in terms of the optical depth. The optical depth corresponding 
to equation~(\ref{bri_tem}) is $\simeq 10^{-6}$. 

We now discuss the feasibility of the detection of the fine structure line towards 
SDSS J1030{+}0524.   We shall assume that the neutral
fraction, $f_{\rm neu}$, outside the Stromgren sphere of  this QSO is unity, consistent with 
the measured GP trough.  The redshifted fine structure line would be expected at 
$\simeq  1.36 \, \rm GHz$.  The observed line `width', considering natural
broadening and the Hubble flow across the Stromgren sphere, is expected to 
be approximately $\Delta \nu \simeq 100 \, \rm MHz/(1+z)$; using $z = 6.28$, we obtain 
$\Delta \nu = 13.7 \, \rm MHz$.  The fine structure line would be expected to originate
in a shell that is roughly the size of the Stromgren sphere, and fall of as 
$1/r^2$ beyond this shell, where $r$ is the distance from the QSO.  In the case of SDSS J1030{+}0524,
the angular size of the Stromgren sphere is expected to be 15~arcmin. 

The frequency of the redshifted line is in the observing bands, and the line width is 
within the spectral line capabilities, of several currently operational telescopes.
However, large-collecting-area arrays like the Giant Metrewave Radio Telescope
(GMRT) have large aperture antennas of 45-m diameter
and, therefore, poor surface brightness sensitivity for such extended structures.
The Australia Telescope Compact Array (ATCA), with 22-m antennas, 
has reasonable brightness sensitivity for this problem and, additionally,
operates with 128-MHz bandwidths in spectral line mode.
At $\nu \simeq 1.3 \, \rm GHz$, the ATCA has a system 
temperature $T_{\rm sys} \simeq 25 \, \rm K$ and antenna efficient $K = 0.1 \, \rm K \, Jy^{-1}$. 
The five movable 22-m diameter antennas may be configured into an ultra-compact  
2-D close-packed H75 (75-m maximum baseline) array, and this would yield a number of baselines
sensitive to the 15-arcmin scale fine-structure line signal.  The brightness sensitivity
of this array, for a 8-arcmin scale structure, is 400~$\mu$K in 6~hr integration time.  
The brightness sensitivity in Fourier synthesis images could be enhanced somewhat, by factors of 
a few, by appropriately weighting the baselines to match the synthetic beam to the expected structure scale;
however, the required integration times for a detection are still in the ball
park of $10^{3}$~hr.

The detection of the signal from cosmological H~{\sc ii} regions, however, 
might be feasible using facilities under construction  or planned for the near future, 
like the xNTD in Australia or the Square Kilometer Array (SKA).  These arrays would have smaller antenna
sizes---making the detection of these large-angular-scale structures detectable 
in interferometers---and significantly more numbers of antennas, giving more numbers of 
short baselines that would
usefully respond to the large-angular-scale fine-structure line.  However, the array
configuration designs would have to factor in the extraordinary brightness temperature sensitivity
requirements for this demanding observation.

As an example, the SKA might consist of about $10^{4}$ 12--15~m class reflector antennas,
with aperture efficiency of 60\%, and the system temperature at 1.4~GHz might be about 20~K.
Assuming that the visibilities are optimally weighted so that the synthetic beam of the
Fourier synthesis array is matched to the source size of about 15~arcmin FWHM, the line
strength would have a peak of about 20~$\mu$Jy.  A 5-$\sigma$ detection of the fine structure line 
towards a typical QSO, in a reasonable integration time of about 1~hr, will require 
that about 250 baselines (just 0.001\% of the total) be within about 50~m.

The signal from the recombination epoch is observable
in the frequency band 6--8~MHz as a broad decrement in the brightness temperature 
of the extragalactic background sky, and would be extremely small compared to the 
orders of magnitude more intense Galactic non-thermal emission as well as the average 
low-frequency background brightness temperature arising from the numerous extragalactic
radio sources.  This decrement may be considered to be a distortion to the CMBR spectrum at 
long wavelengths, and would be an all-sky cosmological signal. 
However, owing to the ionosphere, the frequency range in which this feature is expected to appear
is too low a radio frequency to be easily accessible using
ground-based observatories. Therefore, 
even though the observation of this signal might be yet another tool to probe 
the epoch of recombination, new custom-made instruments, which should presumably 
operate from space and above the ionosphere, will require to be built if a
 detection
of this signal is to be attempted.  An additional cause for concern is that 
the Galactic
and Extragalactic background radiations might have low-frequency spectral
 turnovers at these 
frequencies, as a result of free-free absorption as well as synchrotron 
self-absorption, and
these would result in significant spectral features in the band that would 
require a careful
modelling in order to detect any CMBR decrement feature arising from fine 
structure transition
absorption.  Interference from terrestrial man-made transmitters, as well 
as from 
auroral phenomena and solar system objects would also be an issue.

\section{Summary and Discussion}
We have discussed the possibility of detecting the hitherto undetected fine structure line of 
$2p\hbox{--}2s$ transition in two cosmological settings: the epoch of recombination at $z\simeq 1089$ and 
the epoch of reionization at $z \simeq $6--15. 

The expectations for the line from the environments
of ionization sources in the epoch of reionization are interesting and worthy
 of attention
as a novel tool for the investigation of the reionization process and the 
cosmological evolution of the
gas.  The signal is expected to be observable
as an extended and weakly absorbing source, which causes a decrement in the
 brightness of the 
background CMBR, and may be detected by interferometers as a negative source 
akin to the
Sunyaev-Zeldovich decrements observed along the lines of sight through hot
 gas in clusters
of galaxies.  Detection of the $2p\hbox{--}2s$ line signal from the
 outskirts of 
cosmological H~{\sc ii} regions at different redshifts within the 
reionization era may serve to 
determine the neutral fraction of the medium during 
the epoch of reionization, which is a quantity of significant interest 
in modern 
cosmology.  In particular, we have computed a representative signal 
strength by adopting parameters
typical of a QSO observed at $z \simeq 6$.  These QSOs show GP troughs 
in their spectra;
however, owing to the weakness of the GP test, the spectra have only 
been useful in setting 
weak limits on the neutral fraction (constraining the neutral fraction to be
$\ga 10^{-3}$) outside the observed H~{\sc ii} region.  
We have shown that for a fully neutral medium the line peak may 
reach $\simeq 20 \, \rm \mu K$, which is potentially observable by 
radio interferometers that are being designed today.  

Other interesting probes of the reionization epoch include detecting 
 OI line from this epoch (Hernandez-Monteagudo et~al. 2006). 

It is of interest  to compare the relative difficulty associated with 
detecting the 
neutral hydrogen in this indirect way, using the redshifted fine structure 
line, with direct
imaging of the redshifted 21-cm line from neutral hydrogen during the 
epoch of 
reionization (see, {\it e.g.}, Sethi 2005; Zaldarriaga, Furlanetto, \& Hernquist 2004).  
The `all-sky' H~{\sc i} signal might be detectable with a peak strength of 
$\simeq 50 \, \rm mK$; however, it could be very difficult to detect owing 
to calibration and foreground contamination issues ({\it e.g.}, Zaldarriaga et~al. 2004;
Shaver et~al. 1999,  and references therein). 
A  better approach might be to attempt to detect the fluctuating component
of the sky signal, which could have peak intensities of 
$\simeq \hbox{a few} \, \rm mK$ at observing frequencies of $\nu \simeq 100\hbox{--}200 \, \rm MHz$
(Zaldarriga et~al. 2004; Shaver et~al. 1999). 
This translates to roughly the same 
signal strength (specific intensity) as we have obtained in Eq.~(\ref{bri_tem}) for the fine structure line. This is not entirely unexpected: even though the 
level populations of the excited states are much smaller as compared to 
 the ground level
population needed in computing the HI signal, the A coefficient of the 
fine structure transition we consider here is roughly 8 orders of magnitude 
larger than the HI hyperfine transition A coefficient.  
Currently, there are many ongoing and planned 
radio interferometer experiments for the detection of the redshifted H~{\sc i} emission/absorption
from the epoch of reionization ({\it e.g.}, Pen, Wu, \& Peterson 2004; 
the LOFAR project at {\tt www.lofar.org}; 
the MWA project at {\tt www.haystack.mit.edu/ast/arrays/mwa/site/index.html}).  The detection
of the H~{\sc i} signal is firstly more difficult---requiring greater 
sensitivity---because the typical frequency width of the 
signal $\simeq 0.5 \, \rm MHz$ (Zaldarriga et~al. 2004), which is far smaller than the typical width 
expected for the fine structure line ($\simeq 10 \, \rm MHz$).  Second, the
fine structure line is expected at 1.4~GHz, where the sky background temperatures
are significantly lower than in the $100\hbox{--}200 \, \rm MHz$ band, making the telescope system
temperatures lower.  An additional advantage of the 
the indirect detection is that the redshifted fine structure line appears at higher frequencies
($\ga 1 \, \rm GHz$), which are relativity free of interference as compared to  
the low frequency band of $100\hbox{--}200 \, \rm MHz$.  
The main advantage of the direct detection is that unlike the method
we suggest here, it is independent of the existence of strong Lyman-$\alpha$ 
emitters at high redshifts. Another advantage of direct detection is that it
may be detected `statistically' and such a detection might be achieved with greater
ease than direct `imaging' (e.g Zaldarriaga et~al. 2004; 
Bharadwaj \& Sethi 2001); however, the foreground subtraction problem becomes a severe constraint for a 
statistical detection. To summarize: if strong Lyman-$\alpha$ emitting sources
are present at high redshifts, they would facilitate the indirect 
detection of the neutral hydrogen via enabling the detectability of the fine structure line. 
The imaging issues and problems associated with the detection of this signal 
appears to be less of a challenge as comparison to the direct detection of redshifted H~{\sc i} from those epochs. 

Another astrophysical context in which the fine structure line might be detectable is 
the environments of high redshift galaxies, which are strong Lyman-$\alpha$ emitters.
As equation~(\ref{bri_tem}) shows, the observed signal depends on the neutral fraction outside
the Stromgren sphere.  Therefore, detection of this signal would constitute an alternate probe of the 
neutral fraction of the IGM at large redshifts. 

The aim of this work has been to
examine the detectability of the fine structure line in cosmological contexts, to point out 
the cosmological significance of detections, and spawn work that may refine the modelling
presented herein and improve the case for appropriate design of future
telescopes, which might enable the detection of the fine structure line
towards multiple sources in the reionization era.

\section*{Appendix A: Photon distribution function}
The evolution of the photon distribution function, neglecting the 
effect of scattering off  moving atoms, is (see {\it e.g.} Rybicki \& Dell'Antonio 1994):
\begin{equation}
{\partial n_\nu \over \partial t} - \nu H {\partial n_\nu \over \partial \nu} = 
A_{2p1s} n_{2p} \phi_\nu - c B_\nu n_{1s} \phi_\nu n_\nu.
\end{equation}
Here $B_\nu = 3/(8\pi)c^2/\nu_\alpha^2 A_{2p1s}$ and $H = \dot a /a$. The equation may be 
written as:
\begin{equation}
{\partial n_\nu \over \partial t} - \nu H {\partial n_\nu \over \partial \nu} = -1/\tau (n_\nu -n_\star).
\label{disfun}
\end{equation}
Here $\tau = 1/(cB_\nu \phi_\nu n_{1s})$ and $f_\star = A_{2p1s}n_{2p}/(B_\nu n_{1s} c)$. 
The equation above lends itself to a ready interpretation. If the 
second term on the left hand side (which is owing to the expansion
of the Universe) was absent, the distribution function will approach $f_\star$ on a time
scale $\simeq \tau$, where $\tau \simeq 3 \times n_{1s} \, {\rm s}\ll \dot a/a$  (
$n_{1s}$ here has units $\rm cm^{-3}$).  It may be readily verified 
that for the recombination epoch and also
the epoch of reionization, $\tau \ll 1/H$, excepting when the neutral fraction 
of the medium is very small. We work here with the assumption that the neutral fraction is 
always large enough so that $\tau \ll  1/H$. In this case the solution
to Eq.~(\ref{disfun}) may be simplified: to leading order
the distribution function approaches $f_\star$ and the 
first order term (of the order of $\tau H$) represents the slow time variation 
of the distribution function owing  to the expansion of the Universe. In this 
approximation, one may write the solution to Eq.~(\ref{disfun}) as:
\begin{equation}
n_\nu \simeq  n_\star - \tau H(t)n_\star.
\label{dissol}
\end{equation} 
Using these equations we may proceed to prove the contention that the net
effect of the `trapping' of Lyman-$\alpha$ photons is to reduce the 
decay time of the $2p$ state by a factor $\tau_{\scriptscriptstyle GP}$. 
The $1s\hbox{--}2p$ transition rate, which is given by Eq.~(\ref{levpop1}),
may be solved using Eq.~(\ref{dissol}):   
\begin{equation}
c n_{1s}\int B_\nu \phi(\nu) n_(\nu) d\nu \simeq A_{2p1s}n_{2p} - A_{2p1s}n_{2p}/\tau_{\scriptscriptstyle GP}.
\end{equation}
The first term on the right hand side  cancels with the decay term 
of the $2p$ state on the right hand side of Eq.~(\ref{levpop1}), and, therefore,
the net effect of the scattering of recombination photons is to reduce the 
decay time of the $2p$ state by a factor $\tau_{\scriptscriptstyle GP}$. 
It may be pointed out that the condition needed to derive the above 
expression roughly translates to the condition that $\tau_{\scriptscriptstyle GP} \gg 1$. 
For the reionization case, this requires that the neutral fraction
$\ga 10^{-5}$. In the case of primordial recombination, it leads to an even weaker
condition that the neutral fraction is $\ga 10^{-7}$. 

\section*{Acknowledgment}
One of us (SKS) would like to thank Jens Chluba for many useful discussions
and to Zoltan Haiman for many useful comments on the manuscript.

\newpage

\begin{figure}
\epsfig{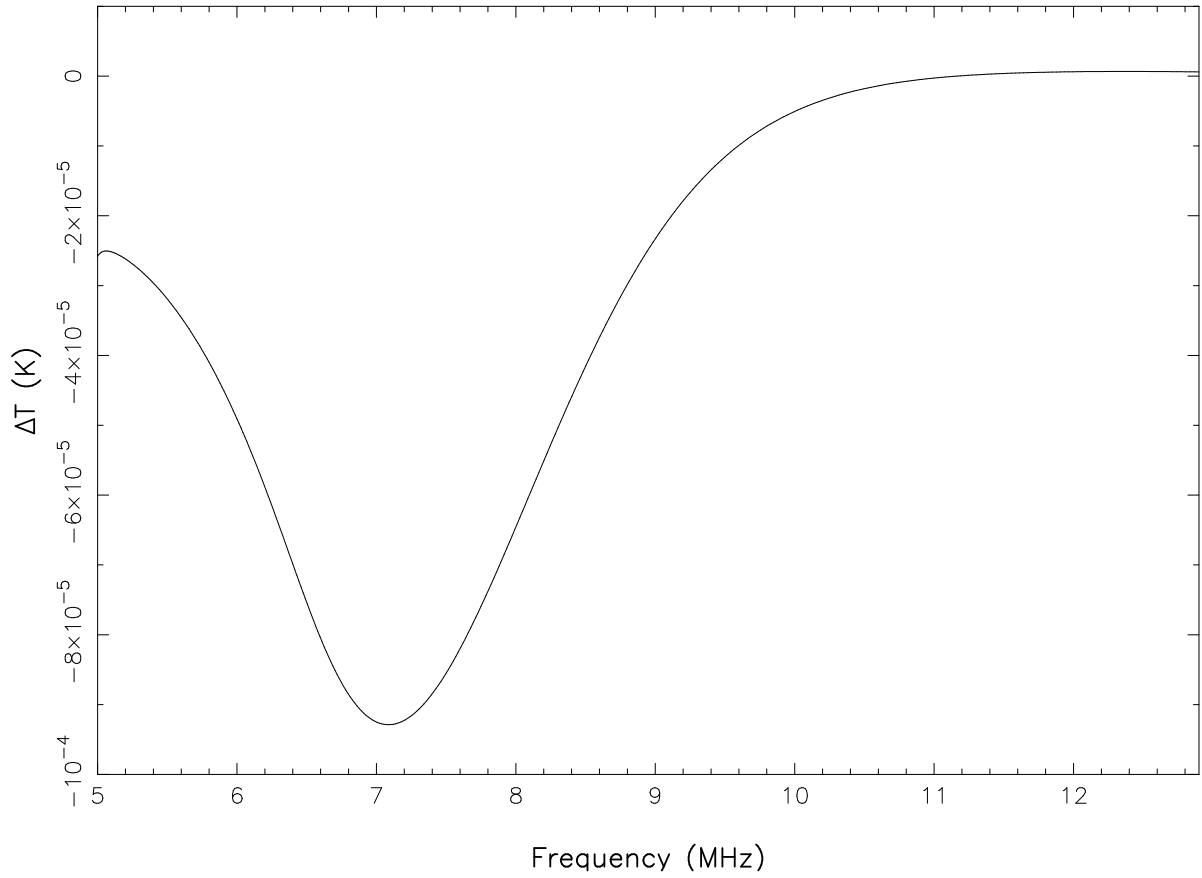}
\caption{The expected brightness temperature decrement in the background radiation, owing to the
fine structure transition in gas at the recombination epoch, is plotted 
versus the observing frequency.}
\label{fig:f1}
\end{figure}

\end{document}